# Hydrogen bonding and percolation in propan-2-ol – water liquid mixtures: X-ray diffraction experiments and computer simulations

*Szilvia Pothoczki\*[1], Ildikó Pethes[1], László Pusztai[1,2], László Temleitner[1], Dániel Csókás[3], Shinji Kohara[4], Koji Ohara[5], and Imre Bakó\*[3]*

[1]Wigner Research Centre for Physics, H-1121 Budapest, Konkoly Thege M. út 29-33., Hungary

[2]International Research Organization for Advanced Science and Technology (IROAST), Kumamoto University, 2-39-1 Kurokami, Chuo-ku, Kumamoto, 860-8555, Japan

[3]Research Centre for Natural Sciences, H-1117 Budapest, Magyar tudósok körútja 2., Hungary

[4] Research Center for Advanced Measurement and Characterization, National Institute for Materials Science (NIMS), 1–1–1 Kouto, Sayo-cho, Sayo-gun, Hyogo 679–5148, Japan

[5]Diffraction and Scattering Division, JASRI, Spring-8, 1-1-1, Kouto, Sayo-cho, Sayo-gun, Hyogo 679-5198, Japan

\*E-mail: pothoczki.szilvia@wigner.mta.hu; Phone: +36 1 392 1469

\*E-mail: bako.imre@ttk.mta.hu; Phone: +36 1 382 6981

**Abstract**

Synchrotron X-ray diffraction measurements have been conducted on aqueous mixtures of propan-2-ol (a.k.a. isopropanol, or 2-propanol), for alcohol contents between 10 and 90 molar %, from room temperature down to 230 K. Molecular dynamics simulations, by using an all-atom parametrization of the propan-2-ol molecule and the well-known TIP4P/2005 water model, were able to provide semi-quantitative descriptions of the measured total structure factors. Various quantities related to hydrogen bonding, like hydrogen bond numbers, size distribution of cyclic entities and cluster size distributions, have been determined from the particle co-ordinates obtained from the simulations. The percolation threshold at room temperature could be estimated to be between isopropanol concentrations of 62 and 74 molar %, whereas at very low temperature, calculations yielded a value above 90 molar %.

# 1. Introduction

The presence of hydroxyl groups (OH groups) in macromolecules, sugars, alcohols, as well as in water, has a significant effect on the structure and function of the molecules, and on the interactions with their environment. [1-5] Mixtures of water and small alcohols, even though they are considered as the simplest systems in which competition between hydrophobic and hydrophilic interactions plays a key role, have a rather complex behavior in many aspects. This is manifested, for example, in their anomalous thermodynamic and transport properties (diffusion coefficient, reorientation correlation time) [6-10]. Despite considerable interest in the field, and the resulting numerous studies, a clear picture has yet to emerge to elucidate satisfactorily these properties. A widely accepted explanation is that these anomalies are connected to perturbations of the H-bond network that is formed by the water and alcohol molecules. [11-16] It is known that properties of this network change significantly as we vary the alcohol concentration and/or decrease the temperature. [17-18]

Quite recently we systematically investigated structural and dynamical aspects of the variations occurring in methanol-, ethanol- and isopropanol-water mixtures as a function of temperature in the water rich region (low alcohol concentration) [19-22]. These studies were mainly focused on the changes in terms of the occurrence of cyclic entities (hydrogen bonded rings) and in terms of the pair interaction energies. It was found that the number of hydrogen bonded rings has increased by lowering the temperature. Six-membered rings were found typical in methanol-water mixtures [19], similarly to pure water. [23] However, for isopropanol- and ethanol-water mixtures the dominance has shifted from six-membered to five-membered rings. [20,22] Furthermore, their water-water interactions become stronger, while solvent-solvent ones appeared to be significantly weaker when compared to the pure substances. [21]

We herein report findings of an extended follow-up study of Ref. [22]. In this contribution, the entire alcohol concentration range was considered, from room temperature down to around the freezing point. For this purpose, new X-ray diffraction experiments were performed at the SPring-8 synchrotron source (Hyogo, Japan). In this way, a complementary collection of diffraction data was provided to Refs. [24-25]. It is worth noting already here that during the present experiments, not even at the lowest temperature, way below the literature value of the freezing point, we were able to detect any hint of crystal formation, such as Bragg peaks: most probably, amorphous phases have been formed. Another, more theoretical, novelty of the present work is a comprehensive percolation analysis. It was performed with the aim of determining the (temperature dependence of the) percolation threshold, by taking into account both (1) the adjacency matrix of H-bonded network, and (2) the Cartesian coordinates of oxygen atoms which form the network. The analyses are based on recent results of network science [26-34].

The paper is organised as follows: in Section 2, details of the diffraction measurements, while in Section 3, those of the computational methods are provided. Section 4 presents results and their discussion, along with an in-depth analysis of the percolation threshold of the hydrogen-bonded clusters. Finally, Section 5 summarizes our findings.

## 2. Synchrotron X-ray diffraction experiments

Samples of propan-2-ol-water mixtures have been prepared for synchrotron X-ray diffraction experiments with alcohol contents of 10.006 (ca. 10), 16.002 (ca. 16), 23.985 (ca 24), 36.013 (ca. 36), 45.034 (ca. 45), 56.065 (ca. 56), 74.035 (ca. 74), 90.192 (ca. 90) and 100 molar % of propan-2-ol. The experiments were performed at the BL04B2 [35] high energy X-ray diffraction beamline of the Japan Synchrotron Radiation Research Institute (SPring-8, Hyogo, Japan). Radiation wavelength was 0.2029 Å, that corresponds to the photon energy of 61.117 keV. Diffraction patterns of the liquid samples were taken in transmission mode, in the horizontal scattering plane, between scattering variable, $Q$, values of 0.16 and 16 Å$^{-1}$, using an array of 6 point detectors. Samples were contained in 2.7 mm inner diameter quartz capillaries, with a wall thickness of 0.1 mm. Three capillaries could be mounted in the automatic sample changer that is connected directly to the cold head of a closed circle refrigerator (CCR). The entire sample environment was under vacuum during the experiments. Diffraction patterns have been recorded starting from room temperature down to 253K for the 10% and 16%, to 210K for the 56%, to 180 K for the 90% and 100% molar %, and to 230K for the remaining samples. In the cases of the 90 and 100% mixtures, no crystallization, indicated by Bragg-peaks, could be observed even at the lowest achievable temperature of the CCR (approx. 15K). Measured raw intensities were normalized by the incoming beam monitor counts, corrected for absorption, polarization and contributions from the empty capillary. The patterns over the entire Q-range were obtained by removing inelastic (Compton) scattering contributions following a standard procedure [36]. The obtained total scattering structure factors are shown in the Supplementary Material, Figures S1 to S5.

## 3. Calculation details

3.1 Molecular Dynamics Simulations

Molecular Dynamics (MD) simulations were performed by using the Gromacs software [37] (version 5.1.1) for various concentration—temperature value pairs of isopropanol-water mixtures; these are plotted by black solid squares in Figure 1. For 2-propanol molecules, the all-atom optimized potentials for liquid simulations (OPLS-AA) [38] force field was used. Bond lengths were kept fixed by the LINCS algorithm [39]. Parameters of atom types and atomic charges can be found in the Supplementary material of Ref [22].

Figure 1. Phase diagram of isopropanol-water mixtures [40]. Gray area: solid state; white area: liquid state (as determined experimentally). Black solid squares: present MD simulations; green solid triangles: new X-ray diffraction data sets; red crossed empty circle: X-ray diffraction data sets from Ref. [24]; blue solid squares: our earlier MD simulations from Ref. [22].

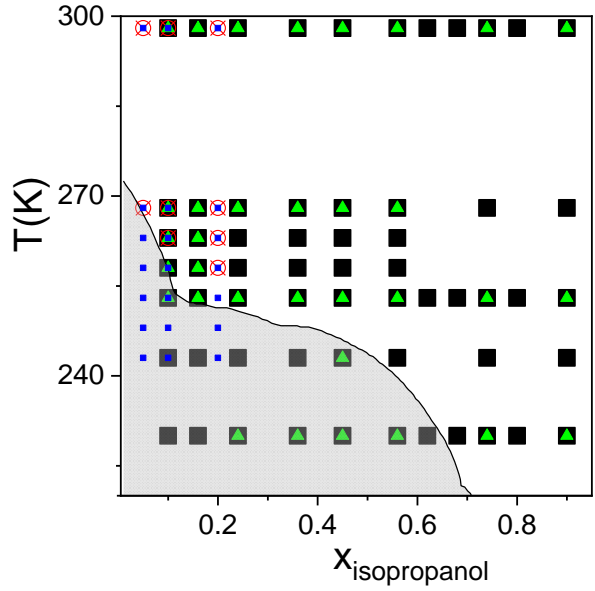

Similarly, as before [20,22], we tested two different (SPC/E [41] and TIP4P/2005 [42]) water models, whether one of them can provide a superior fit to the experimental structure factors. Based on results shown in Section 4.1 systems with the TIP4P/2005 water model were chosen for subsequent analyses (see below). Simulation box lengths and bulk densities can be found in Table 1. In each case 4000 molecules (with respect to compositions and densities) were placed in a cubic box, with periodic boundary conditions.

Table 1: Box lengths, together with the corresponding bulk densities, for each simulated system. (The composition-temperature pairs connected to the newly obtained X-ray diffraction data sets are highlighted in bold.)

| $x_{ip}$ | | 298 K | 268 K | 263 K | 258 K | 253 K | 243 K | 230 K |
|---|---|---|---|---|---|---|---|---|
| 0.1 | L(nm) | **5.35075** | **5.31637** | **5.30332** | **5.30026** | **5.30129** | 5.29084 | 5.28257 |
| | $\rho$ (g/cm$^3$) | **0.9636** | **0.9824** | **0.9897** | **0.9914** | **0.9908** | 0.9967 | 1.0014 |
| 0.16 | L(nm) | **5.58145** | **5.54067** | **5.53778** | **5.52701** | **5.52338** | 5.50705 | 5.49341 |
| | $\rho$ (g/cm$^3$) | **0.9454** | **0.9665** | **0.9701** | **0.9737** | **0.9756** | 0.9843 | 0.9916 |
| 0.24 | L(nm) | **5.87772** | **5.81716** | 5.80791 | 5.8088 | **5.79486** | 5.77648 | **5.75398** |
| | $\rho$ (g/cm$^3$) | **0.9197** | **0.9487** | 0.9533 | 0.9528 | **0.9597** | 0.9689 | **0.9803** |
| 0.36 | L(nm) | **6.27998** | **6.20831** | 6.20346 | 6.19374 | **6.18502** | 6.15757 | **6.14195** |
| | $\rho$ (g/cm$^3$) | **0.8895** | **0.9206** | 0.9228 | 0.9272 | **0.9311** | 0.9436 | **0.9508** |
| 0.45 | L(nm) | **6.55584** | **6.48633** | 6.46654 | 6.4559 | **6.44955** | 6.42509 | **6.4014** |
| | $\rho$ (g/cm$^3$) | **0.8711** | **0.8994** | 0.9077 | 0.9122 | **0.9149** | 0.9254 | **0.9357** |
| 0.56 | L(nm) | **6.86456** | **6.77936** | 6.76719 | 6.76207 | **6.74684** | 6.72518 | **6.68986** |
| | $\rho$ (g/cm$^3$) | **0.8539** | **0.8865** | 0.8913 | 0.8933 | **0.8994** | 0.9081 | **0.9203** |
| 0.62 | L(nm) | 7.01738 | — | — | — | 6.90684 | — | 6.84716 |
| | $\rho$ (g/cm$^3$) | 0.8478 | | | | 0.8892 | | 0.9126 |
| 0.68 | L(nm) | 7.17343 | — | — | — | 7.04895 | — | 6.99399 |
| | $\rho$ (g/cm$^3$) | 0.8391 | | | | 0.8844 | | 0.9054 |
| 0.74 | L(nm) | **7.3175** | 7.2361 | — | — | **7.17935** | 7.16674 | **7.14201** |
|     | ρ (g/cm³)    | **0.8333** | 0.8618 |     |     | **0.8824** | 0.8871 | **0.8963** |
| 0.8 | L(nm)        | 7.45651 | —       | —   | —   | 0.8726  | —       | 0.8929  |
|     | ρ (g/cm³)    | 0.8281  |         |     |     | 7.32755 |         | 7.27134 |
| 0.9 | L(nm)        | **7.68876** | 7.59944 | — | — | **7.55217** | 7.5249 | **7.48453** |
|     | ρ (g/cm³)    | **0.8168** | 0.8459 |     |     | **0.8619** | 0.8713 | **0.8851** |
| 1.0 | L(nm)        | **8.0662** | 7.80995 | — | — | **7.764** | 7.7376 | **7.70423** |
|     | ρ (g/cm³)    | **0.7036** | 0.7752 |     |     | **0.7890** | 0.7971 | **0.8075** |

The Newtonian equations of motions were integrated via the leapfrog algorithm, using a time step of 2 fs. The particle-mesh Ewald algorithm was used for handling long-range electrostatic forces. [43,44] The cut-off radius for non-bonded interactions was set to 1.1 nm.

We followed exactly the same simulation sequence as before [20]: first NPT systems (at each concentration) were heated up to 340 K, using a Nose-Hoover [45,46] thermostat with a time constant of $\tau_T=1.0$, and a Parrinello-Rahman [47] barostat with a time constant of $\tau_p=4.0$, over 5 ns, in order to avoid the aggregation of 2-propanol molecules. After that, a 5 ns NVT equilibration run with a Berendsen [48] thermostat ($\tau_T=0.5$) was applied. This was the starting point for further simulations. By this sequence, it is possible to exploit that the Berendsen method is a fast, first-order approach to equilibrium, whereas the Nose-Hoover thermostat with Parinello-Rahman barostat provides canonical ensembles with correct fluctuation properties. Furthermore, in NVT simulations it is a good practice to perform the equilibration using the Berendsen thermostat with a small value of $\tau_T$, that could be increased later to obtain a stable trajectory in equilibrium. [49]

Accordingly, for every composition, the following four steps were performed to reach the next lower temperature: 1. *NPT_short run* (2ns, Berendsen thermostat with $\tau_T=0.1$, Berendsen barostat with $\tau_p=0.1$). 2. *NPT_long run* (10ns, Nose-Hoover thermostat with $\tau_T=1.0$, Parrinello-Rahman barostat with $\tau_p=4.0$). 3. *NVT_short run* (1ns, Berendsen thermostat with $\tau_T=0.1$). 4. *NVT_long run* (5ns, Berendsen thermostat with $\tau_T=0.5$). All results reported below were calculated from the NVT_long runs.

Throughout this work we intended to focus mostly on models in the liquid phase. Although measurements were carried out also in the temperature region that is below the solid-liquid phase boundary (cf. Figure 1), evidence for the presence of any crystalline phase was not found. In the experiments, we assign this to the formation of solid amorphous phases. On the simulation side, in every case we checked that our system was in the liquid range, and not in an amorphous frozen material, by means of calculating the self-diffusion coefficients. They were determined from mean square deviations (MSD) of centers of mass by Einstein's method, with the help of the *g_msd* software that is also included in GROMACS simulation package. These self-diffusion coefficients as a function of temperature are presented in Table S1-S8. It can be concluded that even at the lowest studied temperatures the atoms are still mobile enough in the simulations. We therefore have to conclude that, not unexpectedly, the experimental freezing point in isopropanol-water mixtures are higher than those observed in the MD computer simulations.

## 4. Results and discussion

### 4.1. Total scattering structure factors

The validation of our computer models is going to be provided on the basis of comparisons with measured total scattering structure factors (TSSF). 'Simulated' TSSF-s were calculated from the simulated partial radial distribution functions by an in-house code. For calculating partial radial distribution functions the *g_rdf* software was used, that is part of the GROMACS software package.

Figure 2 shows calculated (from molecular dynamics) and measured (by X-ray diffraction) total scattering structure factors using two different (SPC/E and TIP4P/2005) water models for the 56 mol% aqueous solution of isopropanol, as a function of temperature. Measured and calculated TSSF-s for all concentrations and temperatures can be found in the Supplementary Material Figs. S1-S5. Simulated systems with both water models follow the trends of the diffraction data at each concentration and temperature. The limitations of our MD models are detectable in three regions: (1) at low concentrations, MD models somewhat overestimate the intensity of the third peak of the diffraction data; (2) at higher concentrations, as the temperature decreases, MD models do not reproduce the doublet-like shape of the second maxima; (3) the quality of the fit somewhat deteriorates between 8 and 12 Å$^{-1}$ (positions depend on the concentration), due (at least partly) to increasing uncertainty in the measured signals. Apart from these small differences, MD simulations provided reasonable agreements for every concentration-temperature value pairs.

Figure 2. Measured and calculated total scattering X-ray weighted structure factors of the 56 mol % isopropanol aqueous solution with SPC/E (left panel) and TIP4P/2005 (right panel) water models as, a function of temperature.

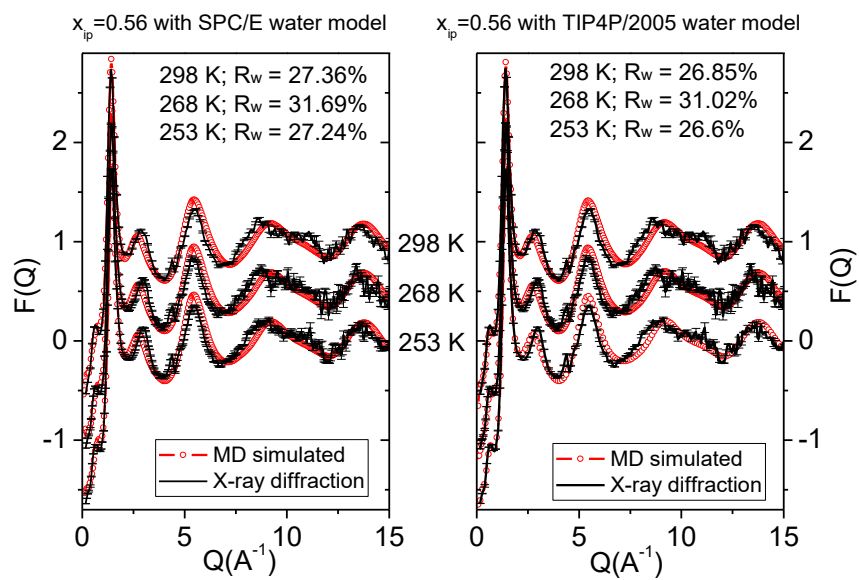

Finally, MD models with the TIP4P/2005 water model were chosen for further analyses, based on the calculated $R_w$ values (cf. Fig. 2 and Figs. S1-S5), which give the difference ($R_W[F(Q)] =$ $\sqrt{\frac{\Sigma_i (F^S(Q_i) - F^E(Q_i))}{\Sigma_i F^E(Q_i)}}$) between MD simulated ($F^S(Q)$) (averaged over many time frames) and experimental structure factors ($F^E(Q)$), thus provide a kind of goodness-of-fit for the models.

Total scattering structure factors of the 10 mol% composition were also measured by Takamuku and co-workers [24] at 298 K, 268 K and 263 K. We compared their measurements with our results, together with the MD models using the TIP4P/2005 water model in Figure 3. We can conclude that the two measurements agree sufficiently well above 1 Å$^{-1}$ at each temperature, considering the (estimated) experimental errors. Slight differences can originate to the different normalization and data processing methods. Below 1 Å$^{-1}$, the new synchrotron data are considered to be more reliable.

Concerning the phase diagram, some studies suggested a liquid-liquid phase separation in the concentration range between 15% and 30%, between 250K and 260K [40, 50], on heating crystallized samples. This was later qualified as metastable phase [51], while other studies [52] could not observe this phase. Based on both the measured total scattering datasets, such behaviour has not been detected.

Figure 3. Total scattering structures factors of the 10 mol% isopropanol-water mixture.

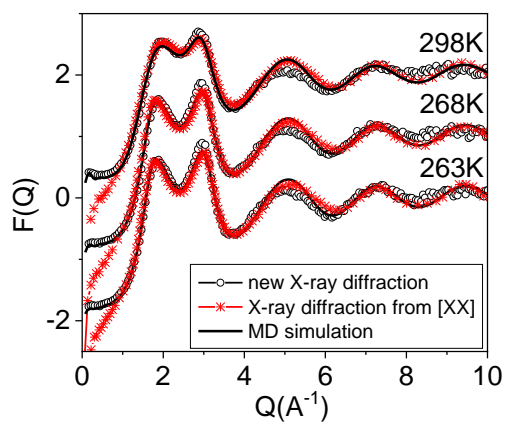

Radial distribution functions are only slightly sensitive to changing temperature in the studied region. Differences are mostly seen between 298 K and temperatures below 273 K, but not between temperatures that are below 273 K. For this reason, these functions are not discussed here in detail; they are shown in the Supplementary material (Figs. S6 to S21).

*4.2. H-bond statistics: average H-bond number, donor and acceptor sites*

In alcohol-water mixtures, where both types of molecules can form hydrogen bonds with themselves and also with each other, H-bonds have a key role for determining the structural and dynamical properties of systems. Following the same definition as in our recent publications [19-21], two molecules have been defined as H-bonded to each other if they are found at a distance r(O···H)

< 2.5 Å, and the interaction energy is smaller than –12 kJ/mol. The two molecules forming a given H-bond can be classified into two groups, according to their roles in the H-bond as either proton acceptors or donors. The following analyses (together with the identification of cyclic entities) were performed using our in-house computer code, described in detail in Ref [53].

The average number of H-bonds, when taking into account connections between all types of molecules, is increasing with the decreasing concentration of isopropanol, as seen in Fig. 4a. Furthermore, on decreasing temperature the number of H-bonded connections between molecules rises significantly. Such behaviour can also be observed for water-water and isopropanol-water H-bonds (c.f. Supp. Fig. S22). With increasing isopropanol content, the number of H-bonds between molecules of propan-2-ol is also increasing, as expected. However, the number of such H-bonds seems to be nearly independent of temperature (Fig. 4b).

Figure 4. Average H-bond numbers, $n_{HB}$, (a) considering each molecule, regardless of their types, and (b) considering isopropanol around isopropanol molecules only.

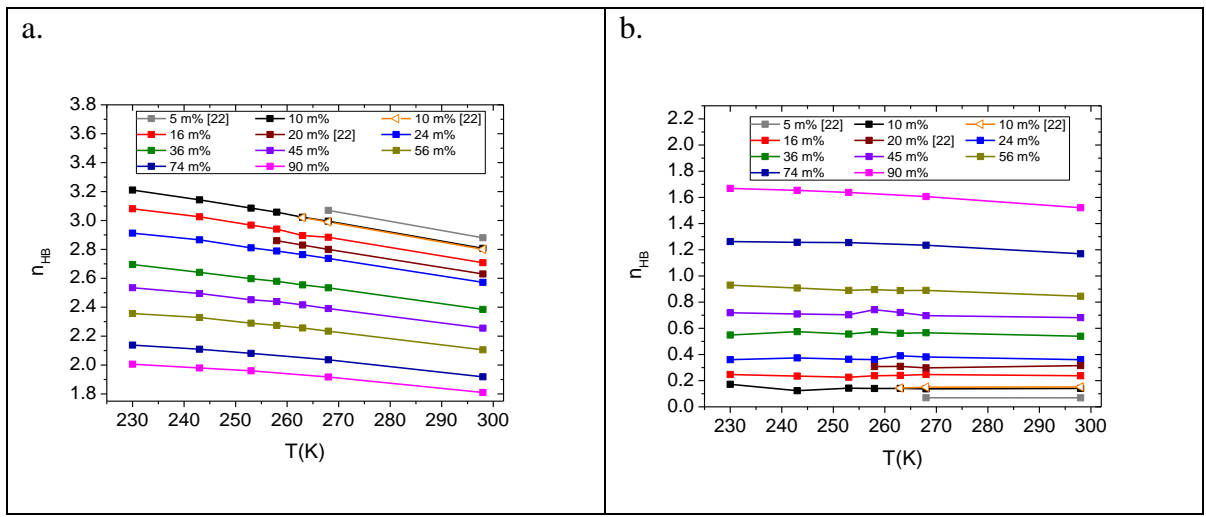

In order to obtain a more detailed picture about the change of coordination shells of water and isopropanol molecules, we calculated the following excess H-bonding numbers:

$$\Delta HB_{wawa} = HB_{wawa} - (1 - x_{ip}) * HB_{wawa\_pure}$$
$$\Delta HB_{ipip} = HB_{ipip} - x_{ip} * HB_{ipip\_pure}$$

where $HB_{wawa}$ and $HB_{ipip}$ are the numbers of H-bonds between two water and two isopropanol molecules in the mixture, $HB_{wawa\_pure}$ and $HB_{ipip\_pure}$ are identical terms in the pure liquids.

The calculated excess H-bonding numbers are plotted in Fig. 5. In the positive value range of the excess H-bond number a well-defined maximum emerges around $x_{ip}$=0.6-0.7 mole fraction for water-water H-bonds. This suggests that at the higher isopropanol concentration range either at lower (230 K) or higher temperature (298 K) water molecules prefer forming H-bonds with themselves. Interestingly, the opposite holds true for isopropanol molecules, that is, they are less likely bonded with other isopropanol molecules. Accordingly, the excess function is negative.

Figure 5. Excess H-bonding numbers.

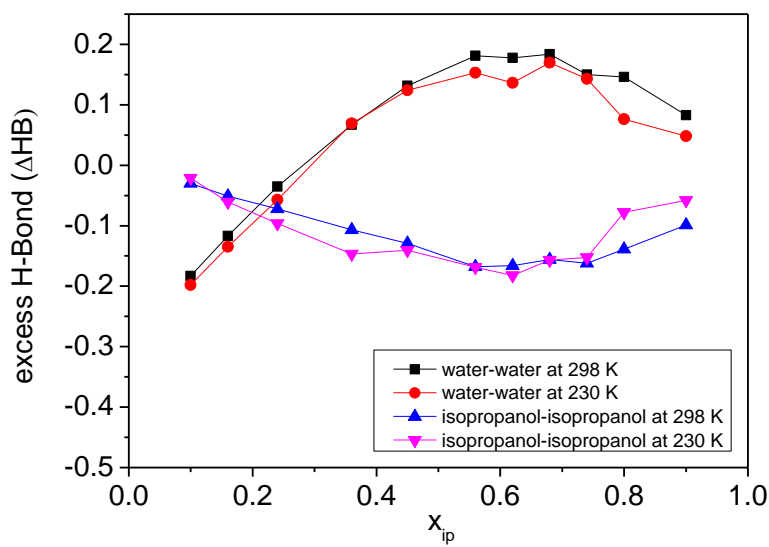

We scrutinized the ratio of donor and acceptor sites of H-bond ('$n_D$ $n_A$') at room temperature as a function of concentration. The number of H-donors ($n_D$) for isopropanol is permitted to take 0 or 1 values, while in the case of water $n_D$=0,1,2. For isopropanol molecules the number of acceptor site ($n_A$) can be 0, 1 and 2, while in the case of water molecules $n_A$=0, 1, 2, 3. As an example molecules were tagged as '0D, 0A' if there is no bond between them.

The results for both the isopropanol and the water molecules are presented in Fig. 6 and Table 2. For isopropanol molecules the number of '1D 1A' group increases upon increasing isopropanol content, while the other three monitored groups ('0D 1A', '2D 1A', '1D 2A') decreases. Up to $x_{ip}$=0.24 the ratio of '0D 1A' and '1D 1A' is very close to 1, while for $x_{ip}$=0.9 the '1D 1A' becomes nearly six times higher than any other groups. By decreasing the temperature also this group ('1D 1A') strengthens at the expense of '1D 1A' and 1D 2A'. This effect is stronger as the isopropanol concentration increases.

Figure 6. Donor and acceptor sites a) at 298 K as a function of isopropanol concentration for isopropanol molecules, b) at 298 K as a function of isopropanol concentration for water molecules.

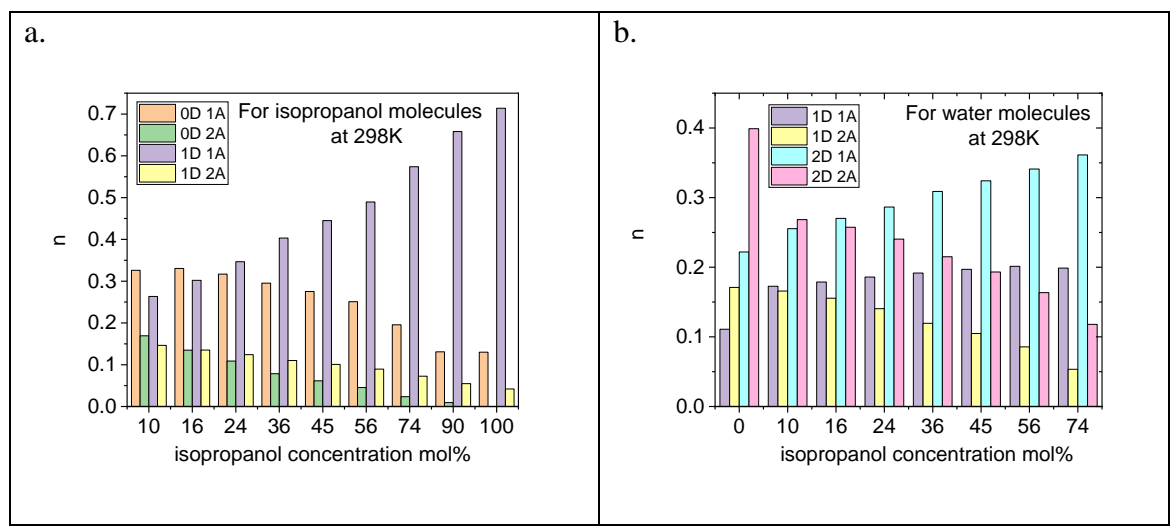

Table 2. Fractions of water and 2-propanol molecules as H-acceptors and as H-donors in the H-bonds identified as a function of temperature and concentration.

|  |  | isopropanol | | | water | | |
| --- | --- | --- | --- | --- | --- | --- | --- |
|  |  | 0D 1A | 1D 1A | 1D 2A | 1D 1A | 2D 1A | 2D 2A |
| 16 mol% | 298K | 0.330 | 0.302 | 0.135 | 0.179 | 0.270 | 0.257 |
|  | 268 K | 0.289 | 0.310 | 0.171 | 0.136 | 0.276 | 0.338 |
|  | 263 K | 0.288 | 0.303 | 0.176 | 0.132 | 0.278 | 0.345 |
|  | 258 K | 0.276 | 0.307 | 0.185 | 0.122 | 0.278 | 0.368 |
|  | 253 K | 0.272 | 0.306 | 0.191 | 0.115 | 0.276 | 0.383 |
|  | 243 K | 0.259 | 0.308 | 0.206 | 0.102 | 0.273 | 0.416 |
|  | 230 K | 0.247 | 0.316 | 0.210 | 0.086 | 0.269 | 0.451 |
| 56 mol% | 298 K | 0.251 | 0.490 | 0.090 | 0.201 | 0.341 | 0.163 |
|  | 268 K | 0.221 | 0.534 | 0.108 | 0.164 | 0.374 | 0.210 |
|  | 263 K | 0.219 | 0.539 | 0.110 | 0.155 | 0.377 | 0.224 |
|  | 258 K | 0.213 | 0.546 | 0.112 | 0.151 | 0.380 | 0.232 |
|  | 253 K | 0.211 | 0.549 | 0.115 | 0.145 | 0.385 | 0.238 |
|  | 243 K | 0.200 | 0.564 | 0.118 | 0.131 | 0.388 | 0.262 |
|  | 230 K | 0.199 | 0.568 | 0.122 | 0.122 | 0.403 | 0.274 |
| 90 mol% | 298K | 0.131 | 0.658 | 0.055 | 0.179 | 0.367 | 0.078 |
|  | 268 K | 0.101 | 0.730 | 0.062 | 0.156 | 0.420 | 0.101 |
|  | 253 K | 0.088 | 0.761 | 0.064 | 0.127 | 0.443 | 0.116 |
|  | 243 K | 0.081 | 0.775 | 0.065 | 0.112 | 0.458 | 0.125 |
|  | 230 K | 0.074 | 0.796 | 0.064 | 0.103 | 0.471 | 0.140 |

Considering water molecules both the '2D 1A' and '2D 2A' cases are prevalent over the water rich region. With increasing isopropanol concentration, the '2D 1A' case stands out while at the same time the '2D 2A' one drops down.

At low isopropanol concentration ($x_{ip} \leq 0.45$) the occurrence of the '2D 2A' case is increasing when the temperature decreases. This is due to the fact molecules tend to maximize their H-bond number. This is in line with our earlier findings. [22] At higher isopropanol concentrations instead of '2D 2A' the '2D 1A' becomes dominant. This '2D 1A' group is also increasing on decreasing temperature.

*4.3 Ring analyses: Cyclic and non-cyclic entities*

Hydrogen-bonded chains can form a continuous path through H-bonds in such a way that the first molecule links to the last one. We refer to these topological units as rings, or primitive rings if the ring cannot be decomposed to smaller rings. In this study, the ring search algorithms developed by Chihaia et al. [53] were used. The following quantities were calculated to characterize the cyclic topology: cyclic size distribution ($n_r$); number of cyclic entities ($N_{cycl}$); number of molecules ($N_{noncycl}$), which are not members of any rings ($n_c < 10$).

Recently, we studied the topology and fine structure of hydrogen bonded aggregations the same way in pure water [23], water-methanol [23], water-ethanol [20] and water-isopropanol mixtures [22], focusing on the water rich region. Our results show that alcohol molecules mostly participate in non-cyclic entities, while water molecules prefer to form rings. That is, the ratio of the components in the rings differs from what we expect based on the stoichiometric composition of the mixture.

First we scanned through the whole concentration range (not only the water rich region as in Ref. [22]) of isopropanol-water mixtures and determined the fraction and number of molecules that participate only in non-cyclic entities. This has been done by considering all molecules, as well as the water and isopropanol subsystems, as a function of composition. These functions are monotonically increasing as can be seen in left hand side of Fig. 7a and b. At low temperature (230 K) the number of non-cyclic entities is significantly smaller than that of at 298 K. This difference is more pronounced at low $x_{ip}$ mole fractions. At both temperatures, for the highest studied isopropanol concentration ($x_{ip}=0.9$), nearly all isopropanol molecules belong to non-cyclic entities. In the right inset of Fig. 7a and b we show that the calculated ratio of noncyclic isopropanol and water molecules, together with their ideal values $x_{ip}/(1-x_{ip})$.

Figure 7. Number of molecules that do not participate in any rings, divided by the total number of molecules, as a function of isopropanol concentration. a) 298 K; b) 230K.

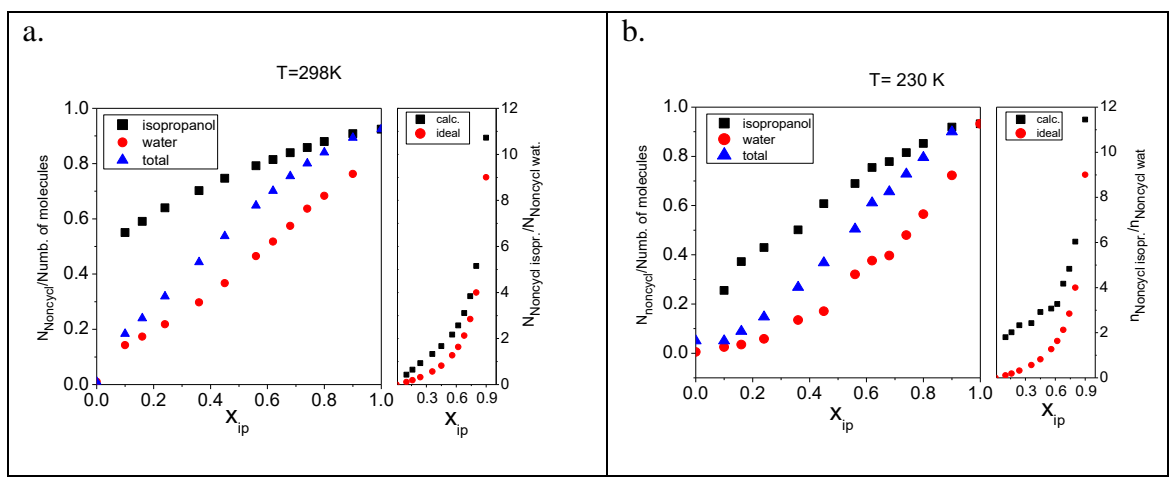

At both temperatures, a well-defined deviation from the ideal behaviour can be found, which is more significant at the lower temperature. This deviation also reflects that the two constituents have different capabilities of forming H-bonds.

It was also observed, both in methanol-water mixtures and in pure water, that the overall number of cycles becomes progressively larger, especially for the 6-membered cycles, as the temperature decreases. [19] On the other hand, for ethanol- and isopropanol-water mixtures, when the alcohol content is low, we found a preference for 5-membered rings, especially at low temperature. Figure 8 (and Fig. S23) reveals that in the present mixtures the number of cycles decreases when the isopropanol concentration, or the temperature increases. The number of cycles in the mixture with the highest studied alcohol concentration ($x_{ip}$=0.9) is not shown, because the number of cyclic entities is very low. For the other concentrations, a preference for 5-membered rings can be observed, just as we have already found at low isopropanol concentration [22]; also note the tendency for forming 4-membered rings at the highest alcohol concentrations and temperatures.

Figure 8. The number of cyclic entities (per particle configuration) as a function of temperature. a) $x_{ip}$=0.1, b) $x_{ip}$=0.16, c) $x_{ip}$=0.56, d) $x_{ip}$=0.74.

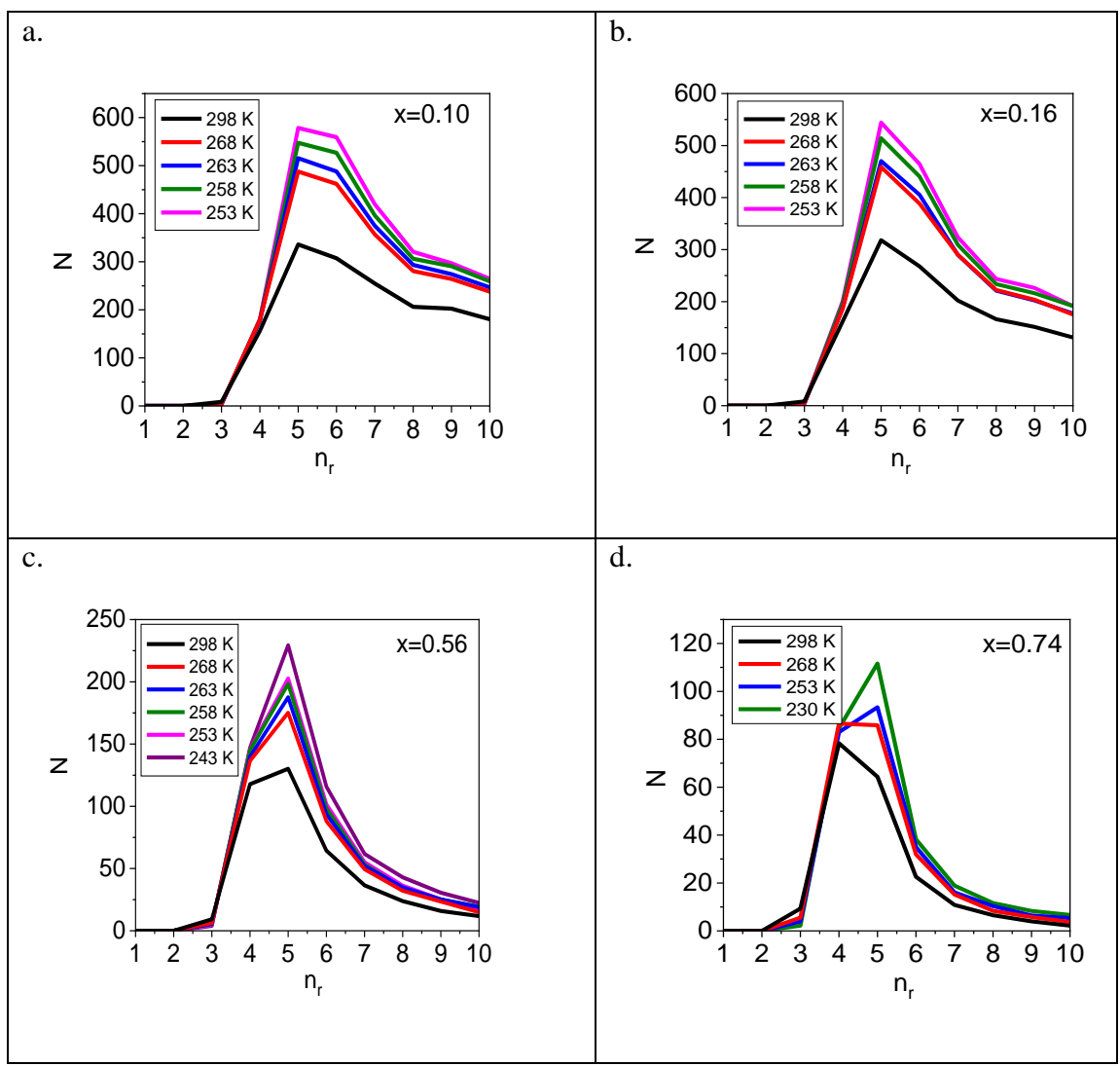

*4.4. Percolation*

Cluster size distributions have been calculated the for each system studied. Two molecules are considered as members of the same cluster if they are connected by a chain of hydrogen bonds. If the largest cluster size is comparable to the system size, then we can say that the connected molecules form a three dimensional percolating hydrogen bonding network. This was shown to be the case for isopropanol-water mixtures in the water rich region ($x_{ip} \leq 0.2$). [22] At higher isopropanol concentrations, isopropanol molecules may disrupt connections between molecules, so that connected clusters may become significantly smaller. A percolation threshold can then be observed, perhaps as a function of both composition and temperature. In this Section, we apply several quantities for characterising the percolation transition in isopropyl alcohol-water systems over the entire concentration range, as a function of temperature.

It is known that if the system is below the percolation transition, the *cluster size distribution*, $P(n_c)$, can be described theoretically by the following formula:

$$P(n_c) = n_c^{-\tau} \exp(-an_c)$$

where $a$ is constant that characterizes a given cluster size distribution, $n_c$ is the cluster size and $\tau$ depends on the dimensionality of the system. [26] Furthermore, the cluster size distribution can be given by $P(n_c)=n_c^{-2.19}$ for random percolation on a 3D cubic lattice. Percolation transition can be ascertained by comparing the calculated cluster size distribution function of the present system with that obtained for the random system. [17,20,27,28]

Figure 9. Cluster size distributions a) at room temperature for different concentrations of isopropanol-water mixtures, b) for $x_{ip}=0.9$ at different temperatures.

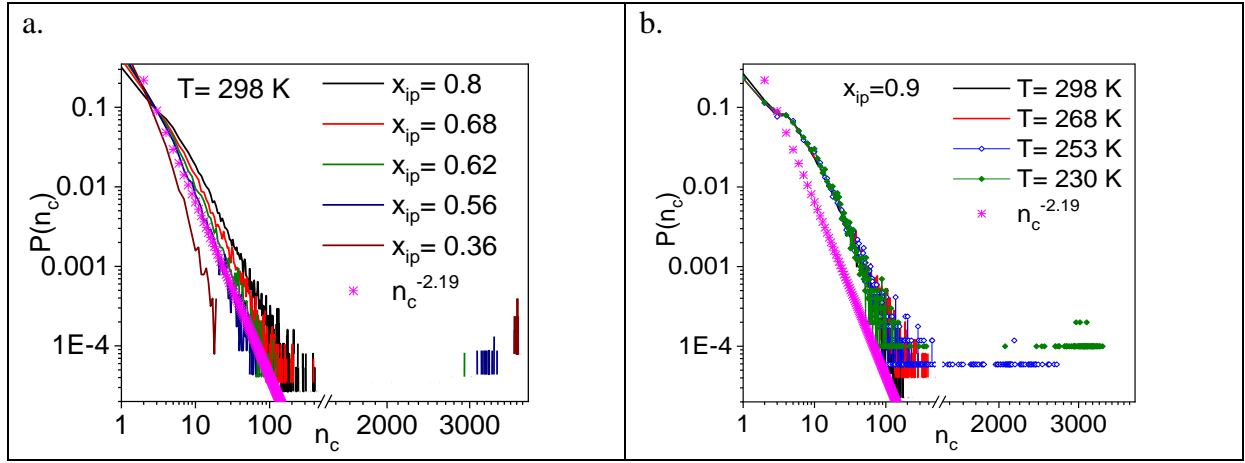

The total H-bonded network at room temperature percolates up to about $x_{ip}\approx 0.56$-$0.62$ mole fraction, as presented in Fig. 9a. At the higher isopropanol concentrations ($x_{ip}>0.62$) with decreasing temperature, it is found that molecules participate in a percolated network only at 230K. This behaviour is demonstrated for the highest isopropanol concentration investigated ($x_{ip}=0.9$) in Fig. 9b. It is worth pointing out that molecules in pure liquid isopropanol do not form an infinite hydrogen bonded network, even at the lowest temperature (c.f. Supp. S24): in the alcohol, only branched chain type structures were detected.

In addition, we calculated the following quantities from the network topology [26,27,29-32], in order to provide the appropriate value of the percolation threshold:

$$Q_1 = \frac{\langle C_2 \rangle \langle C_1^2 \rangle}{\langle C_1 \rangle^2}$$

$$Q_2 = \frac{\langle S_2^2 \rangle}{\langle 3S_2^2 - 2S_4 \rangle}$$

$$S_w = \frac{\langle S_2 \rangle}{\langle S_1 \rangle}$$

where $C_1$ and $C_2$ is the size of the largest and second largest cluster, respectively, $S_1$, $S_2$ and $S_4$ are the first, second and fourth moments of the cluster size distribution. $S_w$ is the cluster size weight average; while calculating, $S_w$, the largest cluster was excluded.

Figure 10. The $<C_2>$, $S_w$, $Q_1$ and $Q_2$ quantities for the characterization of the percolation transition

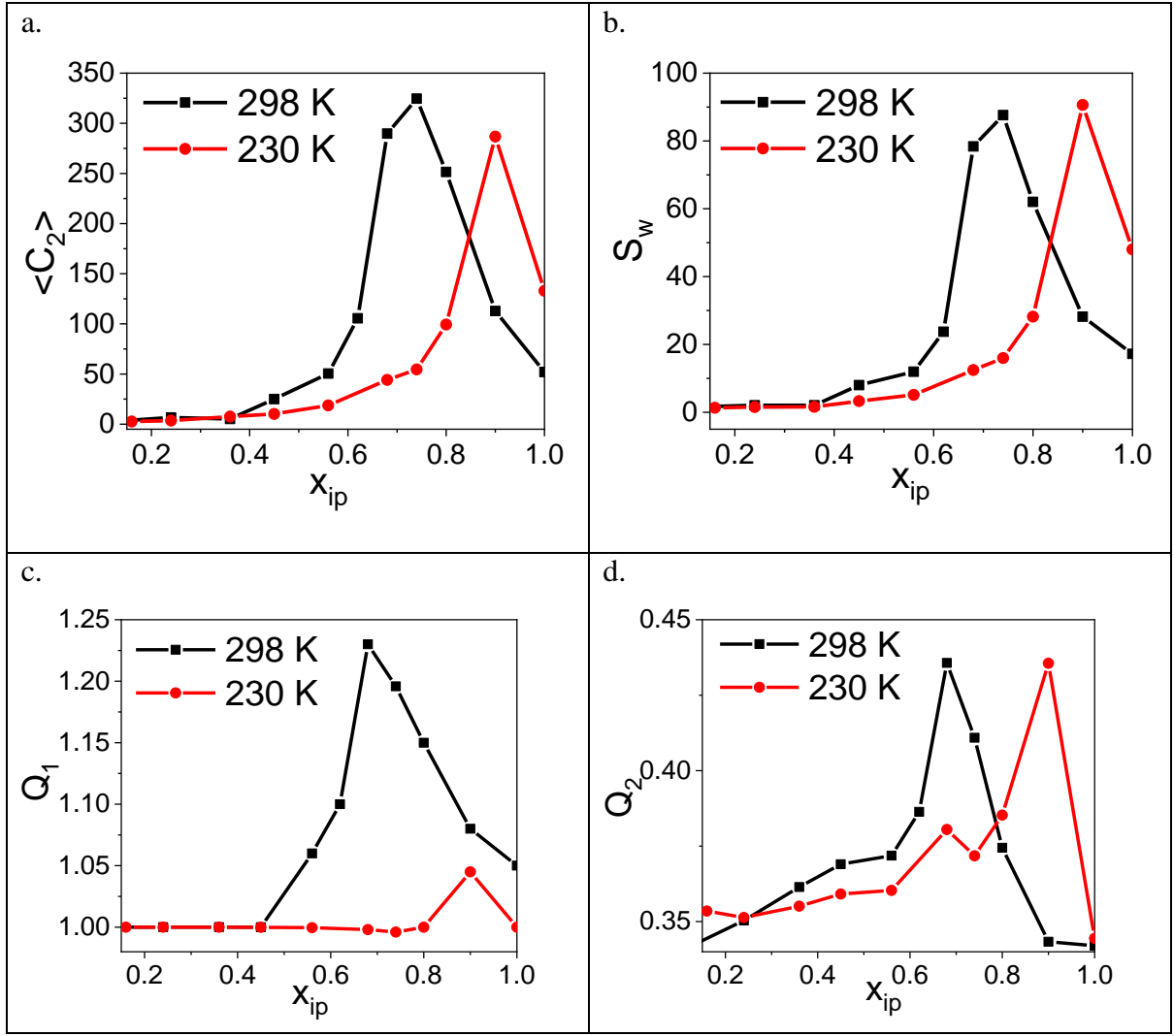

At room temperature all four quantities ($C_2$, $S_w$, $Q_1$, $Q_2$,) have a well-defined maximum around $x_{ip} \approx 0.68$-$0.74$ (Fig. 10). Although a maximum emerges at $x_{ip}=0.9$ for 230 K, we cannot identify unequivocally this concentration as a transition point. We can only estimate that it has to be somewhere there or above it. It is also worth noting that the average largest cluster size ($C_1$) at room temperature around the percolation threshold ($x_{ip} \approx 0.68$-$0.74$) is half of the whole system size (N/2) (c.f. Supp. Mat., Figure S25).

It has already been shown for water and for various aqueous mixtures that the percolation transition can be located by the critical probability ($R_e$) of finding a cluster that spans the system in 3 dimensions. [14,28,29]. Additionally, it has also been proven that this parameter is universal, or at least, depends only very weakly on the system size. [29,30] For this reason, we determined $R_e$ for two different system sizes at 298 K for all concentrations, and in the case of $x_{ip}=0.9$, for all temperatures. The intersection of various system sizes indicates the percolation threshold, which was found to be around $x_{ip}=0.68$ at 298K and $x_{ip}=0.9$ at 253 K. (Fig. 11.)

Figure 11. The critical probability ($R_e$) of finding a cluster that spans the system in 3 dimensions. a) T=298 K b) $x_{ip}=0.9$

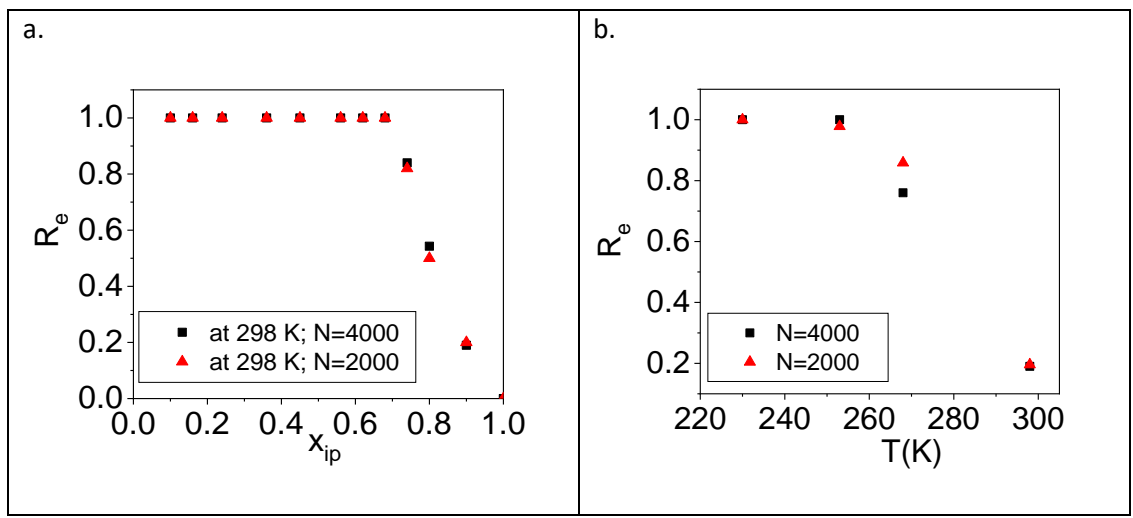

The quantities introduced above intended to describe the network in topological space. We now move on to real space and introduce the fractal dimension of the largest cluster ($d_f$) in the following way:

$$m(r) \approx r^{d_f}$$

where m(r) is the number of molecules that belong to the largest cluster. It has been shown that in three dimensions we cannot detect a percolated cluster with a $d_f$ value smaller than 2.53. [28,32]

Figure 12. Distribution of the dimension of the largest cluster sizes.

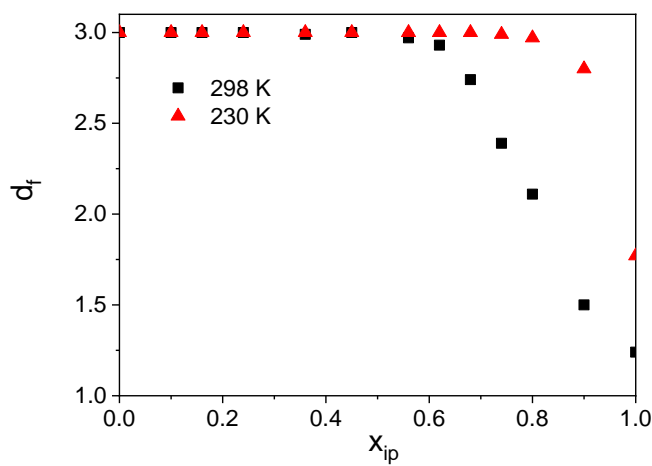

Results displayed in Figure 12 are in line with our findings described above relating to $R_e$. Namely, that a percolation transition at room temperature exists at above $x_{ip}=0.62$, while at 230 K the percolation threshold was found between 0.9 and 1.0 molar fraction.

Typical hydrogen bond network topologies for $x_{ip}=0.1$, 0.62 and 0.9 are shown in Figure 13 as obtained from the molecular dynamics simulation of water-isopropanol mixtures. It is apparent (Fig. 13.a) that at low isopropanol concentration one can find many rings connected to each other over the entire temperature range studied. This result was also found for pure water. [23] When the isopropanol concentration is increased, the system is still percolated, as seen for $x_{ip}=0.62$ at 298 K in Fig. 13.c and 230 K in Fig.13.d, but the number of rings is less than at low isopropanol content.

Concerning the highest isopropanol concentration studied ($x_{ip}=0.9$), at room temperature (Fig. 13.e) mostly branched chain structures (of about 20 to 30 molecules) are observed. On the other hand, at the lowest temperature studied (230 K) the system consists of percolated branched chains and very few rings (Fig. 13.f). It is also worth taking a look at Fig. 13.b, where shorter chains with branching junctions are present in pure isopropanol.

Figure 13. Typical hydrogen-bond topologies a) $x_{ip}=0.1$ at 230 K, b) pure isopropanol at 230 K, c) $x_{ip}=0.62$ at 298 K, d) $x_{ip}=0.62$ at 230 K, e) $x_{ip}=0.9$ at 298 K, f) $x_{ip}=0.9$ at 230 K.

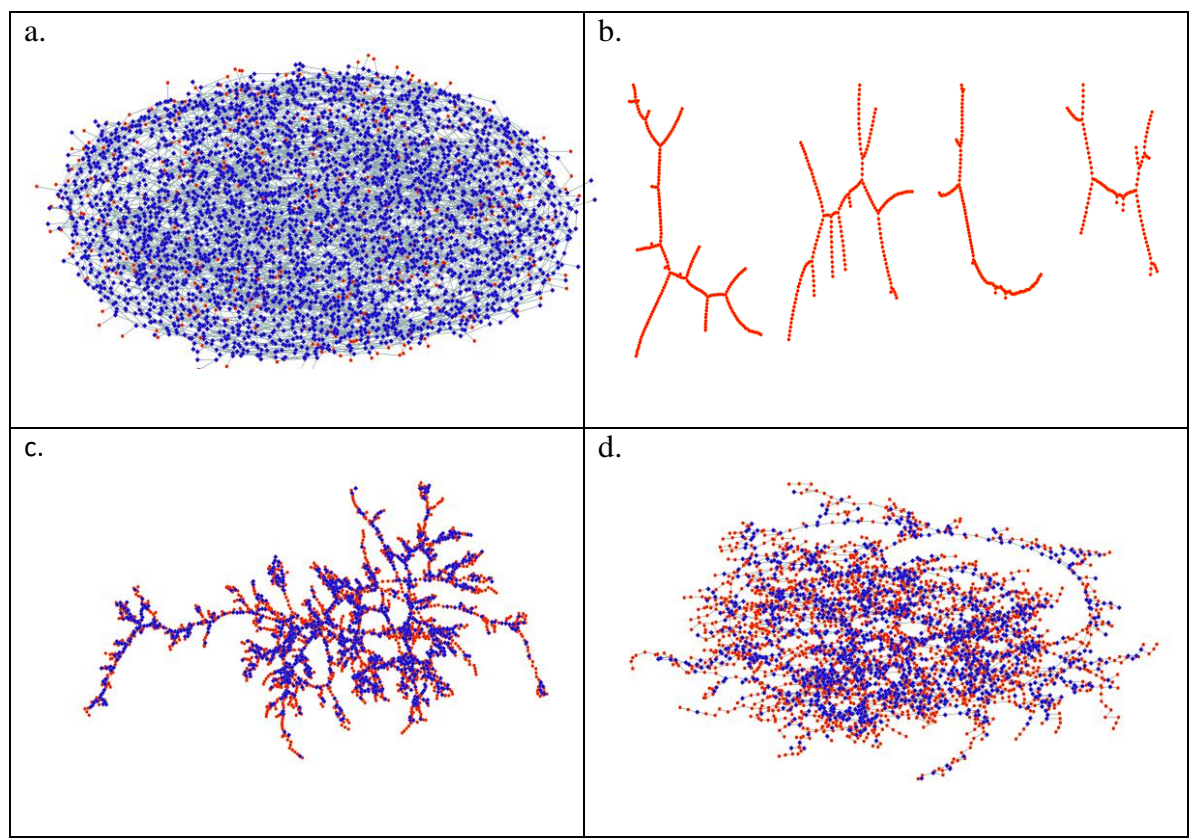

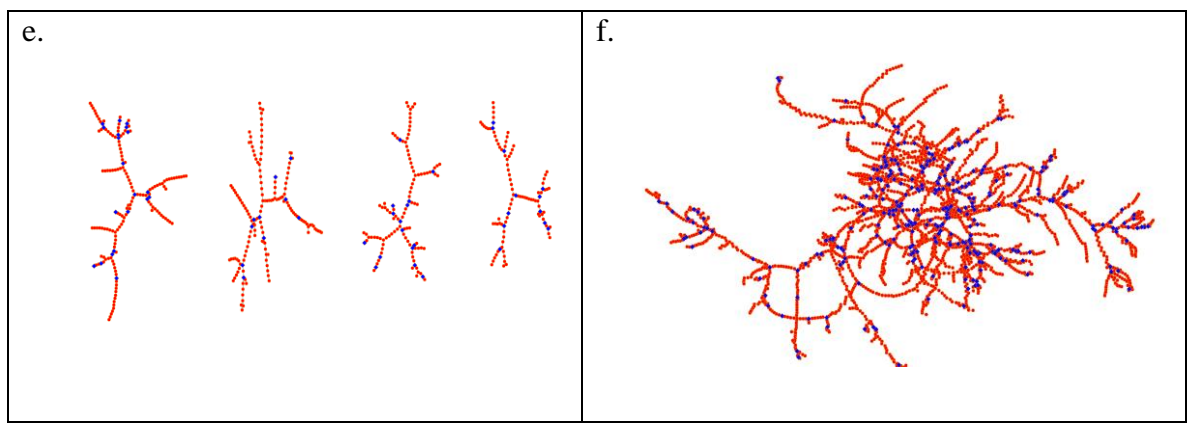

## 5. Conclusions

We have performed new temperature dependent synchrotron X-ray diffraction measurements on isopropanol-water mixtures over the composition range $0.1 < x_{ip} < 0.9$. Diffraction data over the entire composition range are now therefore available, together with Ref. [24], as a function of decreasing temperature. For interpreting diffraction data, molecular dynamics computer simulations have been carried out on a temperature grid that was finer than that of the experiments. From this combined study, the following statements can be made:
  (1) Using the OPLS-AA all-atom potential for the alcohol and the TIP4P/2005 water model, semi-quantitative agreement with the measured total scattering structure factors could be achieved over the composition and temperature range covered. On this basis, details of the structure have been calculated and from the MD particle configurations.
  (2) The number of cyclic entities is decreasing with increasing the alcohol concentration. For a given composition, the number of hydrogen bonded rings is increasing when lowering the temperature. The dominant ring is the 5-fold one.
  (3) The percolation threshold has been estimated by various methods. At room temperature, the percolation transition occurs at around $x_{ip}$=0.62-0.74, whereas at 230 K, the threshold is above $x_{ip}$=0.9.

**Acknowledgement**

The authors are grateful to the National Research, Development and Innovation Office (NRDIO (NKFIH), Hungary) for financial support via grants Nos. KH 130425, 124885 and FK 128656. Synchrotron radiation experiments were performed at the BL04B2 beamline of SPring-8 with the approval of the Japan Synchrotron Radiation Research Institute (JASRI) (Proposal No. 2018B1210). IP and LT are grateful to the mobility support of NRDIO, Hungary, via grant Nos. TÉT_16-1-2016-0202. Sz. Pothoczki and L. Temleitner acknowledges that this project was supported by the János Bolyai Research Scholarship of the Hungarian Academy of Sciences.

**Supplementary Material**

Figure S1: Measured and calculated total scattering structure factors for X-ray diffraction for the mixture with 10 mol % (left panel) and 16 mol % (right panel) isopropanol, as a function of temperature.

Figure S2: Measured and calculated total scattering structure factors for X-ray diffraction for the mixture with 24 mol % (left panel) and 36 mol % (right panel) isopropanol, as a function of temperature.

Figure S3: Measured and calculated total scattering structure factors for X-ray diffraction for the mixture with 45 mol % (left panel) and 56 mol % (right panel) isopropanol, as a function of temperature.

Figure S4: Measured and calculated total scattering structure factors for X-ray diffraction for the mixture with 74 mol % (left panel) and 90 mol % (right panel) isopropanol, as a function of temperature.

Figure S5: Measured and calculated total scattering structure factors for X-ray diffraction for pure isopropanol, as a function of temperature.

Table S1: Calculated self-diffusion coefficients of the components in the 10 mol % isopropanol-water mixture as a function of temperature. (Temperatures where measured structure factors are available are shown in bold.)
Table S2: Calculated self-diffusion coefficients of the components in the 16 mol % isopropanol-water mixture as a function of temperature. (Temperatures where measured structure factors are available are shown in bold.)
Table S3: Calculated self-diffusion coefficients of the components in the 24 mol % isopropanol-water mixture as a function of temperature. (Temperatures where measured structure factors are available are shown in bold.)
Table S4: Calculated self-diffusion coefficients of the components in the 36 mol % isopropanol-water mixture as a function of temperature. (Temperatures where measured structure factors are available are shown in bold.)
Table S5: Calculated self-diffusion coefficients of the components in the 45 mol % isopropanol-water mixture as a function of temperature. (Temperatures where measured structure factors are available are shown in bold.)
Table S6: Calculated self-diffusion coefficients of the components in the 56 mol % isopropanol-water mixture as a function of temperature. (Temperatures where measured structure factors are available are shown in bold.)
Table S7: Calculated self-diffusion coefficients of the components in the 74 mol % isopropanol-water mixture as a function of temperature. (Temperatures where measured structure factors are available are shown in bold.)
Table S8: Calculated self-diffusion coefficients of the components in the 90 mol % isopropanol-water mixture as a function of temperature. (Temperatures where measured structure factors are available are shown in bold.)
Figure S6: Heavy-atom related partial radial distribution functions for the mixture with 10 mol % isopropanol, as a function of temperature.
Figure S7: Heavy-atom related partial radial distribution functions for the mixture with 16 mol % isopropanol as a function of temperature.
Figure S8: Heavy-atom related partial radial distribution functions for the mixture with 24 mol % isopropanol as a function of temperature.
Figure S9: Heavy-atom related partial radial distribution functions for the mixture with 36 mol % isopropanol as a function of temperature.
Figure S10: Heavy-atom related partial radial distribution functions for the mixture with 45 mol % isopropanol as a function of temperature.
Figure S11: Heavy-atom related partial radial distribution functions for the mixture with 56 mol % isopropanol as a function of temperature.
Figure S12: Heavy-atom related partial radial distribution functions for the mixture with 74 mol % isopropanol as a function of temperature.
Figure S13: Heavy-atom related partial radial distribution functions for the mixture with 90 mol % isopropanol as a function of temperature.
Figure S14: H-bond related partial radial distribution functions for the mixture with 10 mol % isopropanol as a function of temperature.
Figure S15: H-bond related partial radial distribution functions for the mixture with 16 mol % isopropanol as a function of temperature.
Figure S16: H-bond related partial radial distribution functions for the mixture with 24 mol % isopropanol as a function of temperature.
Figure S17: H-bond related partial radial distribution functions for the mixture with 36 mol % isopropanol as a function of temperature.
Figure S18: H-bond related partial radial distribution functions for the mixture with 45 mol % isopropanol as a function of temperature.

Figure S19: H-bond related partial radial distribution functions for the mixture with 56 mol % isopropanol as a function of temperature.
Figure S20: H-bond related partial radial distribution functions for the mixture with 74 mol % isopropanol as a function of temperature.
Figure S21: H-bond related partial radial distribution functions for the mixture with 90 mol % isopropanol as a function of temperature.
Figure S22: Total average H-bond and isopropanol-isopropanol H-bond number. a) water-water, b) water-isopropanol.
Figure S23: Cyclic entities as a function of temperature. a) $x_{ip}$=0.24, b) $x_{ip}$=0.36, c) $x_{ip}$=0.45.
Figure S24: Cluster size distributions in pure isopropanol.
Figure S25: Number of molecules in the largest cluster as a function of isopropanol concentration at 298 K.